\documentclass[pra,preprint,showpacs,endfloats]{revtex4}
\usepackage{graphicx,psfrag}

\begin{document}
\title{
Superfluidity and Swallowtails; Hysteretic Behavior in Bose
Condensates} 
\author{Erich J. Mueller}
\email{emueller@mps.ohio-state.edu}
\affiliation{Department of Physics, The Ohio State
  University, Columbus, Ohio 43210}
\date{draft 2.0: \today}

\begin{abstract}
We present a theory of
hysteretic phenomena in Bose gases, 
using superfluidity in
one dimensional rings and in optical lattices as
primary examples.  
Through this study 
we are able to
give a physical interpretation of swallowtail loops 
recently found by many authors
in the mean-field energy 
structure 
of trapped atomic gases.
These loops are a generic sign of hysteresis, and in the present context
are an indication of superfluidity.  We have also calculated the
rate of decay of metastable current carrying states due to quantum
fluctuations.
%
%
%
\end{abstract}
\maketitle

\section{Introduction}
Quantum degenerate bosonic atoms have proven important
for studying macroscopic quantum phenomena 
(for a review see \cite{leggett}).  The order parameter of
the condensed phase is a macroscopic quantum wavefunction which,
unlike single particle wavefunctions, can be directly probed in an
experiment.  The interplay between this macroscopic wavefunction and
interactions leads to a variety of novel effects, the most well known 
of which is superfluidity.  Here we explore superfluid phenomena in a
dilute atomic gas with short range interactions.  As we will
show, {\em superfluidity is naturally viewed as a hysteretic response to
rotation,}\/ motivating a more general study of hysteresis.

In our study of superfluidity,
we quantify the roles played by interactions, finite size effects, and
impurities in the behavior of a weakly interacting gas of
one dimensional Bosons, showing that persistent currents can exist
when the interactions are strong compared to any
impurity potentials, but weak enough to not produce large phase
fluctuations.  We present a 
detailed discussion of the energy landscape
of such gases, revealing a non-trivial topography.
In the limit of weak interactions we
calculate the
lifetimes of persistent currents.

In addition to gaining insights into superfluidity within a one
dimensional geometry, our 
approach provides
an intuitive understanding of {\em 
swallowtail energy loops}
found in mean-field studies of Bose gases within periodic
potentials \cite{Niu,pethick,carr}. 
We show that such loops are
a generic feature of hysteresis and,
in the case of atoms in a periodic potential, 
the loops are a manifestation
of superfluidity.
We discuss the underlying quantum scaffolding which supports this
mean-field structure, and identify other settings where it can be
observed. 

In section~\ref{genintro} we introduce the basic phenomenon of
hysteresis.  The remainder of the introduction discusses superfluidity
and provides examples
of scenarios in which a Bose condensate will behave hysteretically.
Section~\ref{model} discusses microscopic models for superfluidity in
both a ring shaped geometry and in an optical lattice.  The
remainder of this paper analyzes these models.

\subsection{Generic properties of hysteresis}\label{genintro}
Hysteresis is the phenomenon where the state of a physical system depends
upon its history.
The canonical
example is a ferromagnet, which in zero applied magnetic field has a
spontaneous magnetization, conventionally taken to be
in the $\hat z$ direction.  This
magnetization is robust in that 
it is not significantly changed
applying a small field in the 
$-\hat z$ direction.
However, if a strong enough field is applied, the magnetization can
`flip', and point in the $-\hat z$ direction.  When the applied field
is reduced to zero, the magnetization does not revert to its original
orientation, but remains pointing in the $-\hat z$ direction. 
In this example, and the ones that follow,
we see that the response of the system lags behind the applied stimulus.

In a classical system, hysteresis is conveniently thought about
in analogy to the Landau theory of phase transitions
\cite{landau}.  One considers the property of interest (in this case
the magnetization $M$) to be an order
parameter.  An energy landscape is produced
by calculating the energy
of the system as a function of this order parameter.  The applied
field (here the magnetic field $H$) changes this landscape.  

\begin{figure}
\includegraphics[width=\columnwidth]{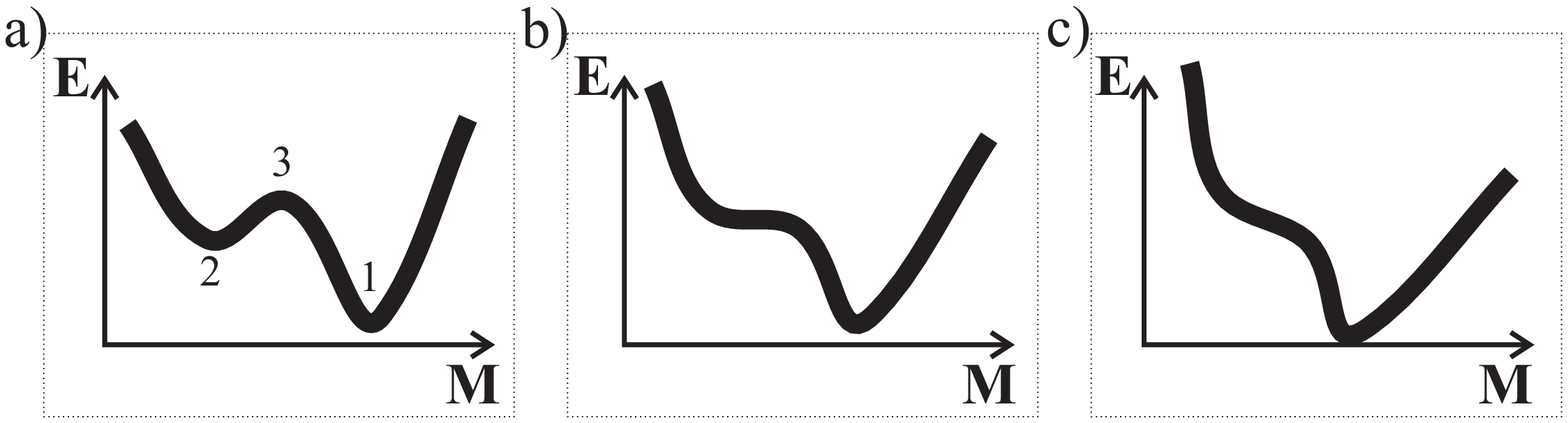}
\caption{Typical energy landscapes: energy $E$ vs order parameter
  $M$.  In a) two minima (labeled 1 and 2) are separated by a
  barrier (3).  In b) one minimum and the barrier coalesce.  In c)
  only one minima exists.  A control parameter ($H$) tunes from one
  landscape to another.  A plot of the energy of the extrema 
  versus the
  control parameter is shown in FIG.~\ref{genericswallow}.
}\label{genlandscape}
\end{figure}

Hysteresis occurs when the energy landscape has more than one minimum,
as depicted in FIG.~\ref{genlandscape}a
(for similar figures
calculated within a microscopic model, see
FIGS.~\ref{barrier} and \ref{landscape}).  For example, both
magnetization in the $\hat z$ and $-\hat z$ directions
might be local minima.
Applying a field tilts the landscape, and reduces the barrier.  At
some critical field, the barrier disappears and the system jumps into
the global minimum (FIG.~\ref{genlandscape}b and c).  
The phenomenon where the barrier disappears goes
under several names; in the theories of phase transitions
\cite{landau} and of gradient dynamics \cite{thom} it is respectively
known as a
spinodal or a
catastrophe.  In more
 mathematical treatments it is referred to
as a ``saddle-node bifurcation''.

\begin{figure}
\includegraphics[width=0.6\columnwidth]{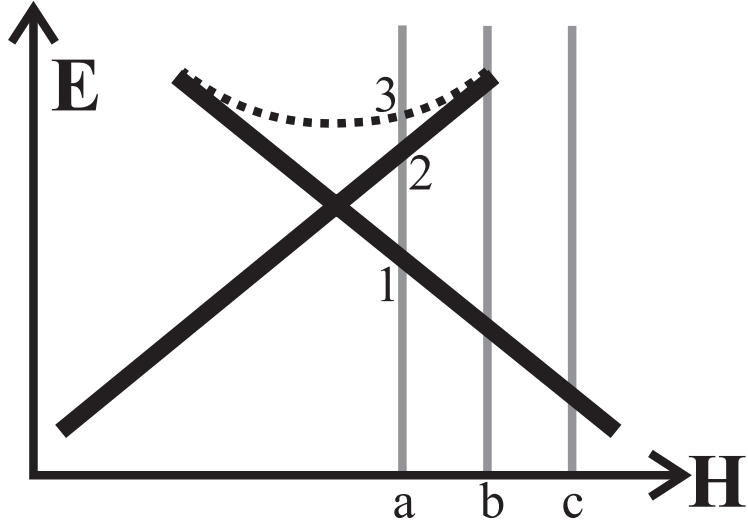}
\caption{Energy extrema as a function of control parameter $H$.
  Solid lines denote minima, dotted denote maxima/saddles.
  The points labeled a, b, and c coincide to the energy landscapes in
  figure~\ref{genlandscape}, which respectively have 3, 2, and 1
  extrema.  The points labeled 1, 2, and 3, coincide with the same
  points in FIG.~\ref{genlandscape}a.
  The existence of multiple minima at the same value of the
  control parameter is a ubiquitous sign of hysteresis.
  The presence of two minima requires a
  maximum/saddle (see FIG.~\ref{genlandscape}).}
\label{genericswallow}
\end{figure}

\begin{figure}
\includegraphics[width=0.6\columnwidth]{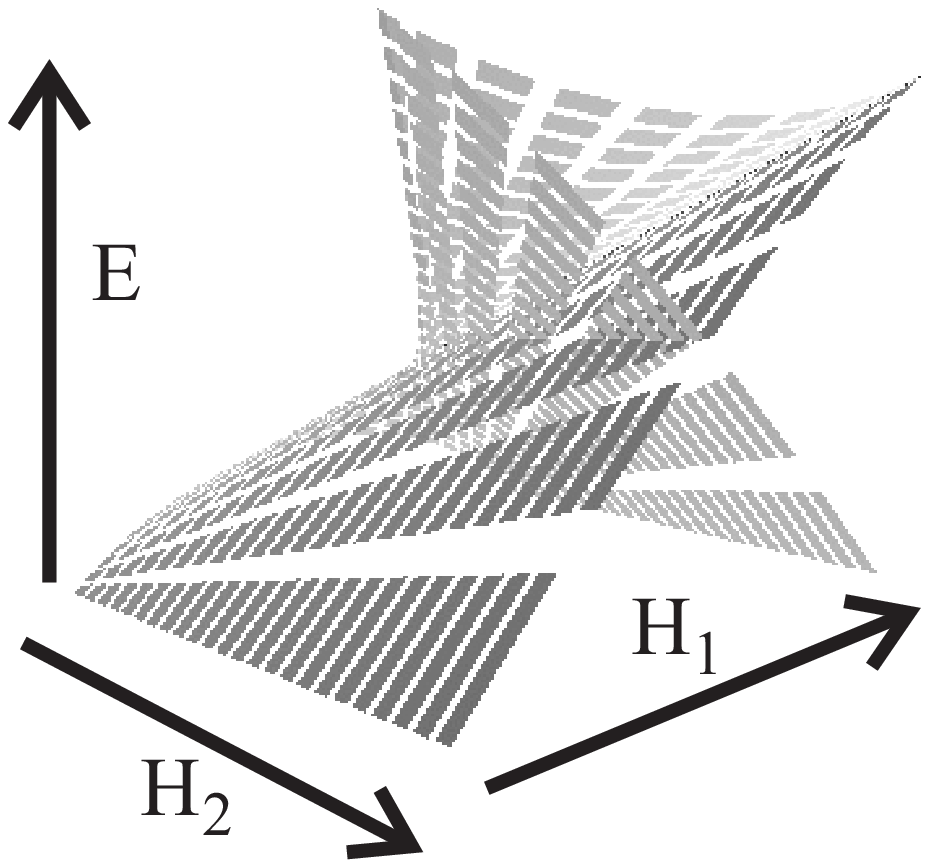}
\caption{Three dimensional depiction of a swallow-tail energy
  structure.  The $x$ and $y$ axes represent control parameters, here
  referred to as $H_1$ and $H_2$, while
  the vertical axis is the energy $E$.  The self-intersecting surface
  shows the stationary points of the energy.  A two-dimensional
  slice is seen in FIG.~\ref{genericswallow} where $H_1$ can be
  identified with $H$.}
\label{threedswallow}
\end{figure}

Figure~\ref{genericswallow} gives a generic
depiction of the energy of the extrema of the
energy landscape (again, similar figures calculated from microscopic
models are shown in FIG.~\ref{energies}).  
A distinctive loop is seen.
This loop, referred to as a ``swallowtail'' by Diakonov et al. 
\cite{pethick}, is a
general feature of the spectrum of a hysteretic system.  It exists
because for some range of fields there are three extrema (two local
minima
and a maximum).  At the point labeled by (c), 
one of the local minima meets up with
the maximum, and they annihilate one another.

To better match the dynamical systems literature, it would be
preferable to not refer to FIG.~\ref{genericswallow} as a
swallowtail, and instead reserve the term for the similar structure
in FIG.~\ref{threedswallow}
that occurs when one has two control parameters.  The second control
parameter changes the size of the loop, and can be tuned so that the
loop, and all hysteresis, vanishes.  
A model which gives rise to this
latter structure will be discussed later.

\begin{figure}
\includegraphics[width=0.6\columnwidth]{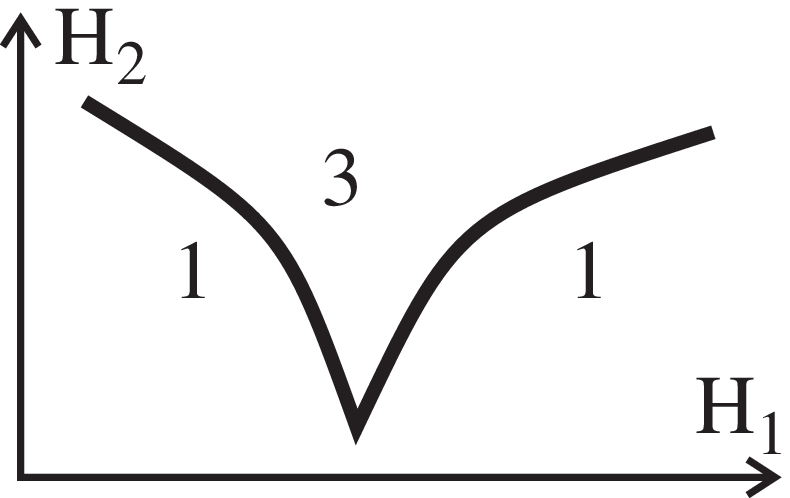}
\caption{Catastrophe set:
values of control parameters for which the number of extrema
  of the energy structure in
  FIG.~\ref{threedswallow} change.  Inside the cusp there are three
  extrema (two minima and a maximum) while outside there is only one.
}\label{cusp} 
\end{figure}

A possible
point of confusion here is that the term ``swallowtail'' is
traditionally used to discuss not the energy structure, but rather the
catastrophe set, which is the points (in the control parameter space)
where  the number of extrema in the energy landscape
change.  The catastrophe set
corresponding to FIGS.~\ref{genericswallow} and \ref{threedswallow}
respectively consists of two points and the cusp-like structure in
FIG.~\ref{cusp}.  Thus the swallow-tail energy spectrum is
associated with a {\em cusp catastrophe} and not a {\em swallow-tail
  catastrophe}. 

The local minima in the energy landscape
are of great physical importance, as the system typically resides
in their vicinity.  Saddle points, and local
maxima, are also important in that the rate of transitions from one
minima to another are governed by the lowest barrier separating the
minima.  In classical systems, these transitions are typically caused
by thermal fluctuations, and occur at a rate proportional to
$e^{-E_b/k_b T}$, where $E_b$ is the barrier height, $k_b$ is
Boltzmann's constant, and $T$ is the temperature.
It should be noted that 
only in very rare physical situations does the system spend
much time at one of these extrema.

In a quantum mechanical system the scenario for hysteresis that we
have discussed becomes more complicated.  The basic
difficulty is that the order parameter is generally not a constant of
motion.  In this case one does not know how to answer questions like
``what is the energy of the system when the magnetization points in
the $+\hat z$ direction?''  There may simply not exist any energy
eigenstates for which the magnetization points in that direction.  
Consequently it is by no means obvious how
to construct an energy landscape, and what significance it will have.

There are three, roughly equivalent, methods of circumventing this
difficulty.  The first approach
is to write the Hamiltonian as a sum of two terms,
$H=H_{\rm diag}+H^\prime$, where the order parameter
commutes with
$H_{\rm diag}$.  This diagonal term is the projection of the Hamiltonian into 
the space where the order parameter has a definite value.
For example, if we have a spin system where
the $z$ component of the
magnetization is our order parameter,
 then $H_{\rm diag}$ would be
diagonal 
in a basis $\{|S,S_z\rangle\}$, where $S$ is the total spin, and $S_z$
is the projection of the spin along the $z$ axis.
If $H^\prime$ is small, one can neglect it
for the sake of drawing the energy landscape.  
The second approach is
to use a variational scheme, where  one writes down ``reasonable''
wavefunctions which are parameterized by the order parameter.  The
expectation value of the energy in these states is an approximation to
the energy landscape.  The final approach is to use a
mean-field theory in which the order parameter is a constant of
motion.  This discussion will be more concrete once microscopic models
are introduced in section~\ref{model} and used to produce energy
landscapes. 

All three of these schemes share the feature that if the system starts
in a local minimum of the energy landscape, there can be matrix
elements in the original Hamiltonian which allow the system to
tunnel to another minimum.  This procedure can be thought of in
analogy to classical thermally activated transport, where due to
thermal fluctuations the system can jump from one minimum to another.
Here it is quantum fluctuations which allow the system to move between
minima.

\subsection{Superfluidity}\label{sfintro}
We now turn to a discussion of superfluidity,
a phenomenon which manifests itself in many related ways, including, 
dissipationless flow,
quantized vortices, reductions in the moment of inertia, and the
existence of persistent currents.  We focus on the latter
phenomenon, which was first observed in $^4$He \cite{hepersist}.  In
an idealized version of 
these experiments, an annular container of helium is rotated while
cooling to below the lambda temperature, where it becomes a
superfluid.  When the container is then stopped, one observes that the
fluid continues to rotate -- maintaining its velocity for extremely
long times.
Arguments based solely on Galilean invariance show that this current
carrying state cannot be the ground state of the system, and is
therefore an extremely long lived metastable excited state
\cite{leggett2}.  
It is observed that the
lifetime of these currents decrease with increasing velocity, and there
is a critical velocity $v_c$, above which no persistent currents exist.

\begin{figure}
\centerline{\includegraphics[width=0.6\columnwidth]{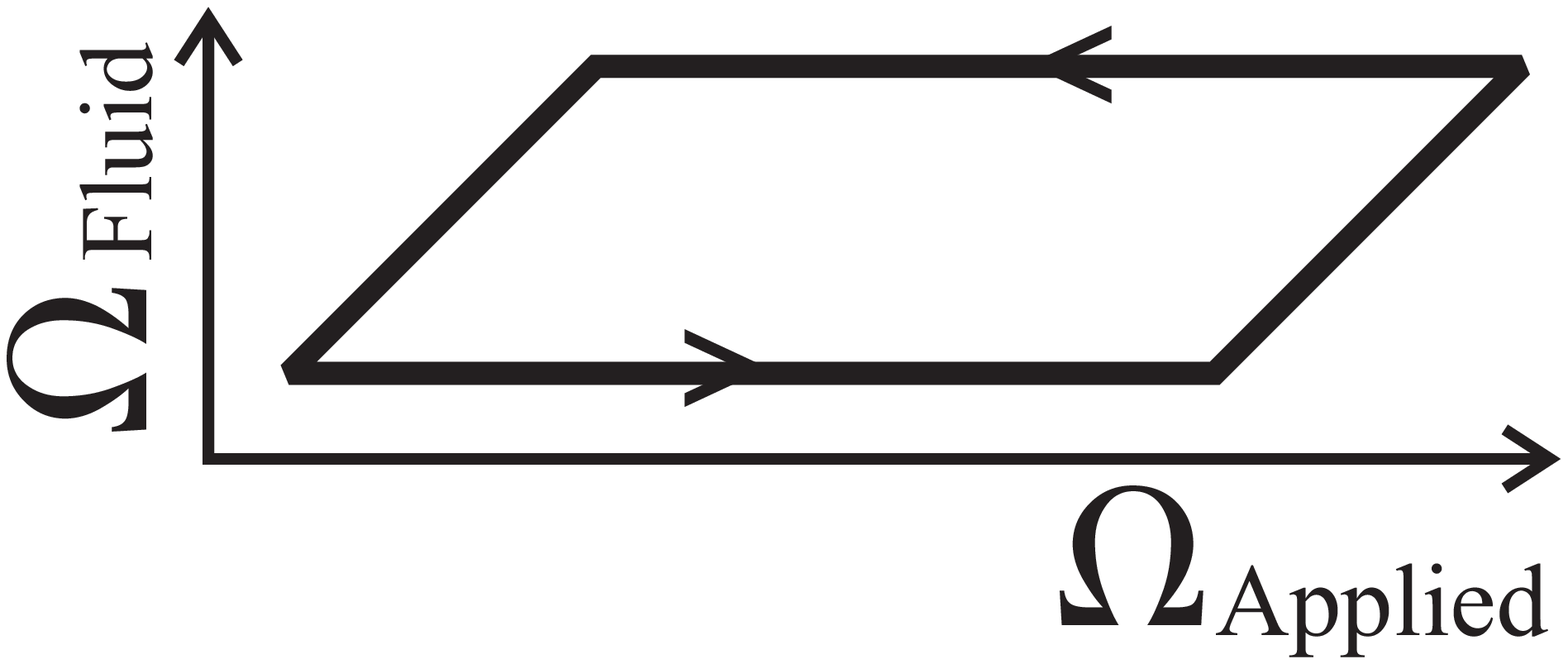}}
\vspace{0.1in}
\caption{Hysteresis loop in a superfluid.  The rotation rate of the
  fluid is shown as a function of the rotation of the container.  The
  arrows denote the direction of the hysteresis loop.  As seen in
  experiments on helium \cite{hepersist}, the sloped lines are
  actually made up of many discrete jumps, which cannot be resolved on
  this scale.}
\label{hyst}
\end{figure}

For our
purposes it is convenient to think of such currents in terms of a
hysteretic response to rotation.  Imagine starting with an annular
container of superfluid which is at rest.  If the container is slowly 
rotated in a clockwise direction the fluid remains at rest (in the
rotating frame this is a persistent current).  If
one rotates faster and faster, the relative velocity between the
container and the fluid eventually exceeds the critical
velocity, excitations are formed, and the fluid accelerates.  At
this point a
persistent current has developed in that even if one stops rotating
the container then the fluid will continue to flow.  
The flow can be stopped 
if one rotates the container sufficiently fast in a counter-clockwise direction.  
The principle is simply that when the relative velocity 
between the fluid and the container exceeds $v_c$, the fluid accelerates.
Thus the fluid flow lags behind the applied rotation, resulting in the hysteresis
loop sketched in FIG.~\ref{hyst}.  By the
arguments of section~\ref{genintro}, one must therefore see energy
structures analogous to those in FIG.~\ref{genlandscape} and
\ref{genericswallow}. 

Here we wish
to understand the origin of this dramatic effect from a microscopic
model.  Standard
descriptions of superfluidity \cite{landau,feynman}
attribute the long life of these currents to the scarcity of low
energy excitations.  In the present setting, it is more natural to
think of superfluidity in terms of the ability of the fluid to screen
out impurities.  The basic argument, which will be given in more
detail later, is that an ordinary (non-super) fluid does not support
persistent currents because the fluid particles scatter off of
small imperfections in the walls of the container, exchanging angular
momentum and eventually equilibrating with those walls.  Through a
collective deformation of the macroscopic wavefunction, quantum
degenerate bosons are able to screen out the imperfections.  Since the
fluid effectively see smooth walls, it does not slow down.  In the
body of this paper these imperfections will be modeled as an
`impurity' potential. 

From this physical picture, one can anticipate many of our results.
In particular, there are two natural control parameters, the rate of
rotation and the strength of interactions relative to the impurity
potential.  We will find 
an energy structure similar to FIG.~\ref{threedswallow}, 
where these two control
parameters correspond to $H_2$ and $H_1$.

\subsection{Optical lattices}\label{latintro}
Superfluidity is not limited to a ring geometry.  As we explain,
superfluid properties naturally appear for Bose particles within a
periodic potential.  Due to their importance in solid-state physics,
quantum phenomena in periodic potentials are very well studied
theoretically and there has been rapid progress on experimental
studies of Bosons in periodic potentials, where the periodicity is
produced using  
standing waves of light (optical lattices)
(for a review see \cite{opticallattices}).
Many of the single particle phenomena of
solid-state physics have been observed in these artificial lattices,
including band structure, Bloch oscillations, and Zener tunneling
\cite{opticallattices}.  
These solid state concepts are reviewed below, and play
an important role in our discussion of superfluidity.
Further theoretical discussions of  these phenomena in
cold gases and relevant discussion of how interactions screen the lattice can be found in 
\cite{niuold}.
Although not directly related to our study of hysteresis, it is worth mentioning that
 correlated many-body states, such as Mott
insulators \cite{mott}, 
have been observed in atoms trapped in an optical lattice.

\begin{figure}
\includegraphics[width=\columnwidth]{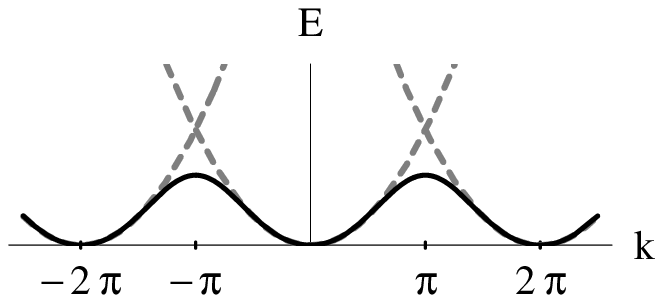}
\caption{Band structure.  The solid line shows the energy of Bloch
  waves of crystal momentum $k$ for non-interacting particles in a periodic
  potential within an extended zone scheme.  The dashed lines show the
  energy states in the absence of the periodic potential.}
\label{bloch}
\end{figure}

Here we use superfluidity to
reexamine theoretical studies of mean-field energy loops in 
of atoms in optical lattices
\cite{Niu,pethick}.
We understand the key features of
these studies 
by
starting with
the energy structure of the non-interacting
single particle states.  As discussed in
textbooks \cite{ashcroft}, the
states available to noninteracting particles in a periodic potential
are labeled by two quantum numbers, a band index $\nu$, and a {\em
  crystal momentum} $k$.  The wavefunctions of these states are of the
Bloch form 
\begin{equation}\label{blochwave}
\psi_{\nu k}(r)=e^{ik\cdot r} v_\nu(r),
\end{equation}
 were $v_\nu(r)$
shares the  periodicity of the lattice, and $k$ is restricted to the
first Brillouin zone.  Limiting our discussion to one dimension with
lattice periodicity $L$, the
first Brillouin zone corresponds to momenta $|k|<\pi/L$.  For
simplicity we use dimensionless units where $L=1$.  In
FIG.~\ref{bloch}, the lowest energy band is sketched in an extended
zone scheme, where the energy is extended periodically to $k$'s
outside of the first Brillouin zone.  This periodicity is the source of
the phenomenon known as
Bloch oscillations.  Imagine starting with a single particle in the
$k=0$ state.  If an external force is applied to the particle, it will
accelerate and $k$ will increase.  For sufficiently weak
acceleration, the state will adiabatically follow the solid curve in
FIG.~\ref{bloch}.  When $k$ has increased to $2\pi$, the system has
returned to its initial state.  Thus a constant force leads to
periodic oscillations.  If the force is too strong, the adiabaticity
condition is violated, transitions are made to higher bands, and one
no longer sees the Bloch oscillations.  This breakdown is known as
Zener tunneling.

\begin{figure}
\includegraphics[width=\columnwidth]{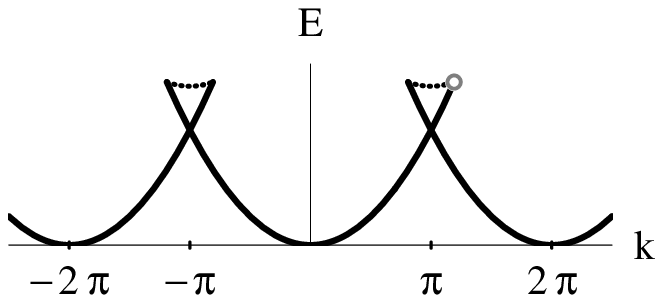}
\caption{Schematic of energy extrema for a condensate in an optical
 lattice.  Solid and dotted lines denote minima and saddle points.
 One of the spinodal points, where the number of extrema change, is
 marked by a grey circle.}\label{sfsw}
\end{figure}

A similar scenario can be considered for
 Bose condensed atoms.  In the ground
state all of the particles reside in the lowest energy Bloch state.
Like the single particle case, when a force is applied, 
the crystal momentum $k$ increases.  However, as a superfluid, the
condensate is able to screen out the periodic potential.  Thus, for
sufficiently strong interactions,
instead of following the solid curve in FIG.~\ref{bloch}, the system
follows a path closer to
the dashed curve, corresponding to the spectrum of states in
the absence of the periodic potential.  The microscopic model which
will be introduced in
 \ref{olmodel} confirms this picture, and one can identify the set of
 states visited during this adiabatic acceleration as local minima in
 a mean-field energy landscape.
When the fluid's velocity exceeds the critical velocity, it looses the
ability to screen the lattice.  Thus the energy curves terminate at
some point, and energy extrema take on the structure in
FIG.~\ref{sfsw}, where one has a crossing of local minima.  
One minimum corresponds to the fluid moving to the right, the other to
fluid moving to the left.  These two states have different momentum,
but share the same crystal momentum.
For purely topological reasons, the presence of two
local minima at a given value of $k$ guarantees that there is a saddle
point separating them.  This barrier state is also shown in
FIG.~\ref{sfsw} as the dotted line forming the `top of the swallow's
tail'.  As will be discussed in SEC.~\ref{sec:mf}, the barrier state
corresponds to a `phase slip'.

The question of what happens to the system when the cloud 
is accelerated past
the `end of the loop' (marked by a grey circle on the figure)
is discussed by Wu et al.
\cite{Niu,niulz}.  Clearly adiabaticity must break down
at this point, and crossing this point from left to right, then back
again will not return the system to its original state.  Thus some
sort of hysteresis has developed.

Note that as the interaction strength is reduced the loops
in FIG.~\ref{sfsw} become smaller and eventually disappear.  Thus if
one identifies $H_1$ with $k$ and $H_2$ with the interaction strength,
the mean-field energy extrema near $k=\pi$ must
have the full swallowtail structure
shown in FIG.~\ref{threedswallow}.

\subsection{Josephson junctions}\label{jjintro}
\begin{figure}
\includegraphics[width=0.5\columnwidth]{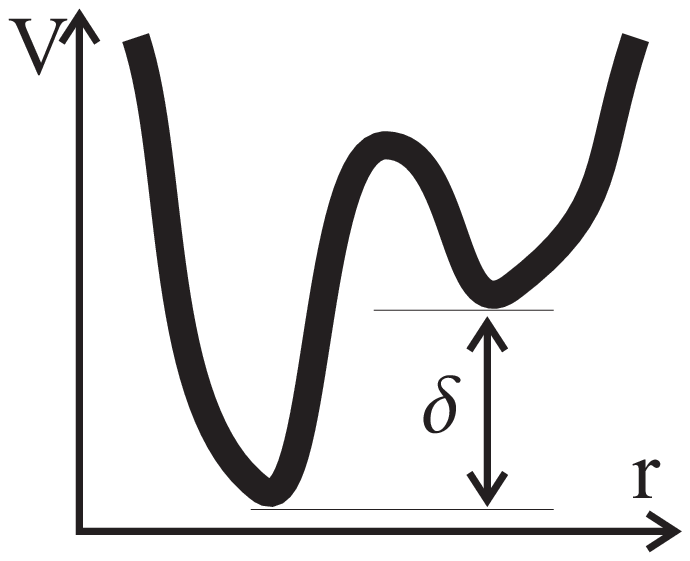}
\caption{Geometry of a double-well trap.  The potential energy $V$ is
  shown as a function of a spatial coordinate $r$.  The two wells have
  an energy difference $\delta$.}\label{dwtrap}
\end{figure}

We conclude the introduction by discussing a hysteretic
Bose system in which the hysteresis is not associated with persistent
currents, namely, a gas of particles with attractive interactions
in a double-well trap as depicted in FIG.~\ref{dwtrap}.  
The control parameters here are the strength of interactions
and the bias $\delta$ which is applied between the
two wells.

\begin{figure}
\includegraphics[width=\columnwidth]{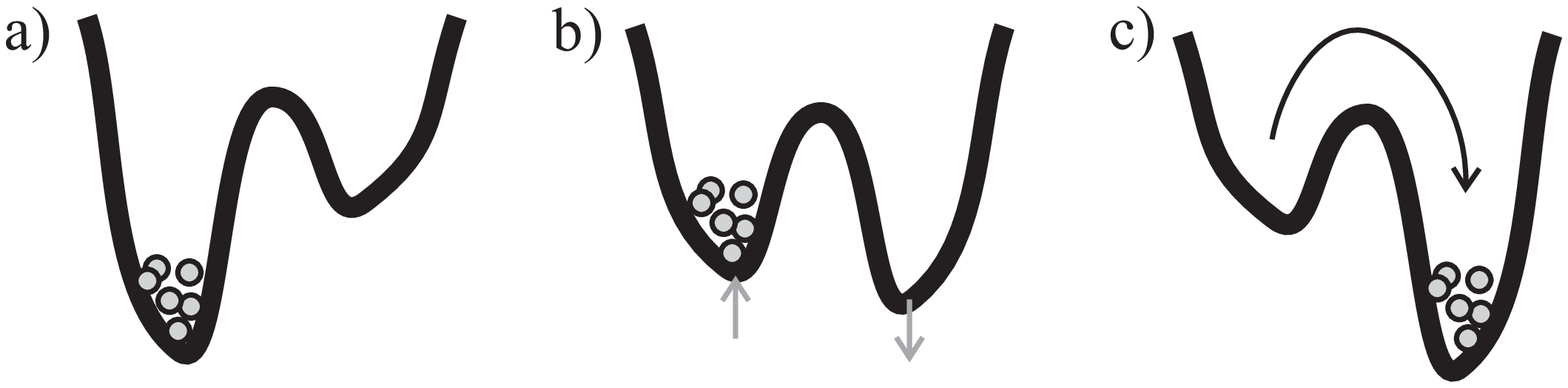}
\caption{Illustration of hysteresis in a double well trap filled with
  attractive bosons.  In each picture the trap from FIG.~\ref{dwtrap}
  is shown with a set value of $\delta$, and the particles shown by
  small grey circles in one of the two wells.  In a) the bias is
  positive, and all of the particles are in the left well.  The bias
  is slowly switched to a small negative value in b).  The particles
  remain in the left hand well, even though the ground state has all
  of them on the right.  For large enough negative detuning c), the
  particles all jump to the right.
}\label{dwhyst}
\end{figure}

FIG.~\ref{dwhyst} illustrates the transformations which give rise to a
hysteresis loop in this system.  One
starts with the left well of much
lower energy than the right ($\delta>0$).  The ground state consists
of all of the particles bunched up on the left.  The bias is then
slowly decreased, and made slightly negative, so that the right hand
well has lower energy.  In the true ground state all of the particles
are sitting in the right hand well.  Nonetheless, 
the particles actually stay in the left hand well.
This behavior is understood by noting that
in order to move the particles from the left hand well to the right
one has to first move a single particle.  Although such a move saves the
potential energy of the bias, separating that one particle from the
others makes the  interaction energy less
negative.  For small enough bias
moving a single particle increases the total
energy, and the state with
all of the particles sitting in the left hand well is a
local minimum of the energy.
If the bias is made more negative, the potential energy savings of
moving a particle to the right hand side eventually becomes greater
than the interaction energy cost.  The particles then all jump to the
right hand well.  The whole process can be reversed, and a hysteresis
loop is formed.  

For weaker interactions, the value of $\delta$ at which the metastable
state disappears becomes smaller.  For sufficiently weak interactions,
no hysteresis occurs.  Thus identifying $H_2$ with the interaction
strength, and $H_1$ with the detuning $\delta$, the energy landscape
has extrema matching FIG.~\ref{threedswallow}.

Experimentally such a double-well trap can be formed by considering
two cells in an optical lattice, or by carefully arranging magnetic and
optical fields as in \cite{mitinterfere}.  We will not explicitly
discuss models for this system, as there exist many excellent
treatments in the literature \cite{jjexample}


\section{Microscopic Models}\label{model}
We now construct microscopic models of the superfluid
systems described in the
introduction.  In addition to verifying the qualitative structures
already discussed, these models allow us to make quantitative
predictions about the behavior of a gas of bosons.  In particular, as
mentioned in section~\ref{genintro}, a quantum mechanical system can
tunnel from one minimum in the energy landscape to another (which
would, for example, lead to the decay of persistent currents).  We are
able to calculate the rate of such tunneling. 

\subsection{Superfluidity}\label{sfmodel}
As a microscopic model of persistent currents, we study a one
dimensional ring of length $L$, rotating at frequency $\Omega$,
containing a cloud of bosons of mass $m$
which interact via short range
interactions.  Measuring energy in terms of $\hbar^2/2mL^2$, ($\hbar$
is Plank's constant) the
Hamiltonian in the rotating frame is
\begin{equation}\label{modham}
H = \sum_j (2\pi j + \Phi)^2 c_j^\dagger c_j 
 + (g/2)\!\!\!\!\!\! \sum_{j+k=l+m}
\!\!\!\!\!\!
c_j^\dagger c_k^\dagger c_l c_m,
\end{equation}
where $\Phi=2 m L^2\Omega/\hbar$ and $g>0$ are dimensionless 
measures of the rotation speed and the strength of the
interactions. Operators $c_j^\dagger$, create bosons with angular
momentum $j\hbar$. 
The model (\ref{modham}) could be experimentally realized by cooling
an atomic gas in an annular trap with harmonic confinement of frequency
$\omega_\perp$ to such an
extent that only the lowest transverse mode is occupied (for
recent experimental progress on annular traps 
see \cite{chapman}).  As long as the trap length
$d_\perp=\sqrt{\hbar/m\omega_\perp}$ is larger than the scattering
  length $a_s$ the interaction parameter would then be given by
  $g=4\pi a_s L/d_\perp^2$.

Despite its apparent simplicity, 
this one dimensional model is quite rich.  It
is a canonical example of a Luttinger liquid \cite{haldane} whose
behavior can be studied via the Bethe ansatz \cite{lieb}.
Two properties worth noting are: 1)
at $g=0$ it describes a non-interacting Bose gas; and 
2) at $g\to\infty$ it can be mapped onto a gas of non-interacting
fermions.  In neither of these limits is the system superfluid, 
however we show that for
small positive $g$ the system {\em is} superfluid. Here we will
study how this superfluidity develops as $g$ is tuned from 0, finding
the structure discussed in section~\ref{sfintro}.  The
equally interesting question of how this superfluidity breaks down as
$g\to\infty$ will not be discussed.

The model Hamiltonian (\ref{modham}) is invariant under rotation, and
therefore conserves angular momentum.  A trivial consequence is that
if a current is started in this system it will never decay. Thus,
as aptly pointed out by Kagan et al. \cite{kagan}, to study superfluidity
one must add an impurity which breaks the symmetry.  In an
experimental setting such terms are always present due to
imperfections in the apparatus.  It is quite instructive to imagine
artificially introducing such an impurity (for instance by using a
laser which interacts with the atoms via dipole forces), and being
able to control its strength.  Conventional discussions of
superfluidity focus on $^4$He, which is strongly interacting, and
whose behavior is largely insensitive to the strength of the
impurities.  In a weakly interacting setting (especially in 1-D)
this is no longer the
case, and the strength of the symmetry breaking term is extremely
important.  The system's behavior is relatively insensitive to the
exact form of the impurity.
Two natural models are a point scatterer $H_{\rm pnt} =
\lambda \sum_{kq} c_k^\dagger c_q$ and a sinusoidal potential 
$H_{\rm sin} =\lambda \sum_k 
\left(c_k^\dagger c_{k-1}+c_k^\dagger c_{k+1}\right)$.  In both cases
$\lambda$ measures the strength of the perturbation.

\subsection{Optical lattices}\label{olmodel}
A model for particles in a periodic potential can be constructed which
has the same structure as (\ref{modham}) with an impurity.  The
rotation speed $\Phi$ and the impurity potential are
respectively mapped onto the crystal momentum and the lattice potential.

Introducing the field operator $\psi(x)$, which annihilates a particle
at position $x$, the Hamiltonian for particles in one dimension
interacting with a local interaction is 
\begin{eqnarray}\label{realspace}
H&=&\int\!\!dx\, \frac{\hbar^2 \nabla \psi^\dagger\cdot\nabla\psi}{2m}+ V(x)
\psi^\dagger \psi + \frac{\bar g}{2}\psi^\dagger\psi^\dagger\psi\psi,
\end{eqnarray}
where $V(x)$ is the lattice potential, $\bar g$ parameterizes the
interactions, and
the argument $x$ is assumed for each of the field operators.  The
periodic potential can be written as $V(x)=\sum_j e^{ip_jx}V_j$, where
$p_j= j k_L= 2\pi j/L$ are reciprocal lattice vectors.

In analogy to (\ref{blochwave}), it is convenient to write our field
operators in a Bloch form,
\begin{equation}
\psi(x) = \sum_j e^{i p_j x} \int_{-\pi}^{\pi} \frac{dk}{\sqrt{2\pi
    L}}  e^{ik x/L} \psi_j(k),
\end{equation}
where $\psi_j(k)$ is the Bose operator which annihilates a particle
with momentum $\hbar (k+2\pi j)/L$.  These obey the
standard commutation
relationships
$[\psi_j(k),\psi^\dagger_{j^\prime}(k^\prime)]=
\delta_{jj^\prime} \delta(k-k^\prime)$. 
In this decomposition, $k$ is the (dimensionless) crystal momentum
which runs from $-\pi$ to $\pi$ and $p_j$ are reciprocal lattice vectors.  The index $j$
plays the role of a band index in the limit that the lattice potential vanishes.

In terms of these operators, the Hamiltonian takes the form
\begin{eqnarray}\label{htot}
H&=& \frac{\hbar^2}{2mL^2}\left(H_{\rm kin}+H_{\rm pot}+H_{\rm
    int}^{\rm (vert)}+H_{\rm int}^\prime\right)\\
\label{hkin}
H_{\rm kin} &=&\textstyle \int\!\! dk \sum_j (2\pi j+k)^2 \psi_j^\dagger \psi_j\\
\label{hpot}
H_{\rm pot} &=&\textstyle \int\!\! dk \sum_{jq} V_{q}  \psi_{j+q}^\dagger \psi_j^\prime\\
\label{hint}
H_{\rm int}^{\rm (vert)} &=&
\textstyle
\int\!\! dk\,(g/2) \sum_{j_1+j_2=j_3+j_4} \psi_{j_1}^\dagger
\psi_{j_2}^\dagger
\psi_{j_3} \psi_{j_4}\\
\label{hint2}
H_{\rm int}^\prime &=&\textstyle
\int\!\!d\{k\}
 (g/2) \sum_{\{j\}} \psi_1^\dagger
\psi_2^\dagger
\psi_3 \psi_4
\end{eqnarray}
where the respective terms in (\ref{htot}) represent kinetic, potential, and
interaction energy.  The interaction is split into two terms, one
$H_{\rm int}^{\rm (vert)}$ only involves particles with the same
crystal momentum, while $H_{\rm int}^\prime$ involves particles with
different crystal momentum.  In (\ref{hkin}) through (\ref{hint}) the
argument $k$ in $\psi_j(k)$ is omitted.  In (\ref{hint2}) the sum and
integral are taken over all $k_i,j_i$ such that momentum is conserved 
\begin{equation}
k_1+k_2-k_3-k_4+2\pi (j_1+j_2-j_3-j_4)=0,
\end{equation}
and where not all of the $k_i$ are equal.  In (\ref{hint2}) the
shorthand notation $\psi_i=\psi_{j_i}(k_i)$ is used.  The interaction
is given by $g=\bar g/2\pi L$.

\begin{figure}
\begin{tabular}{cc}
\parbox[b]{0.04\columnwidth}{a)\\[1in]}
&\includegraphics[width=0.8\columnwidth]{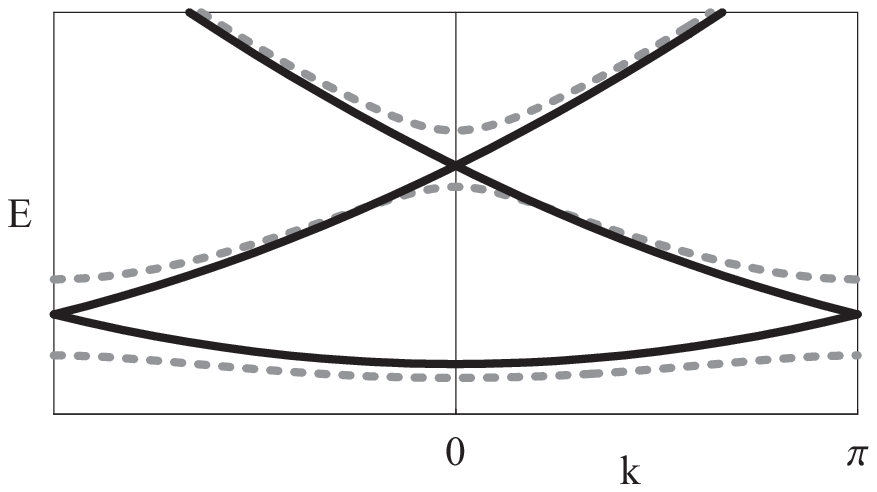}\\
\parbox[b]{0.04\columnwidth}{b)\\[1in]}&
\includegraphics[width=0.8\columnwidth]{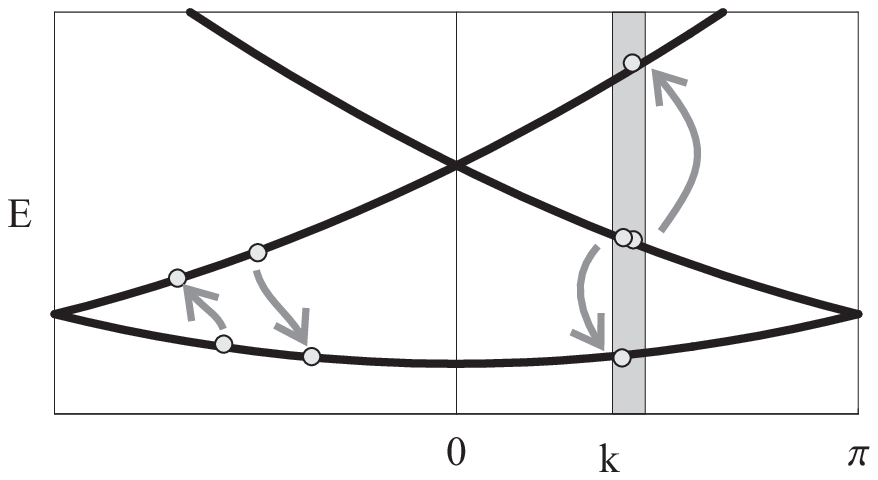}
\end{tabular}
\caption{Illustration of the terms in Hamiltonian (\ref{htot}).  In
  a), the black line shows the kinetic energy $H_{\rm kin}$
  of single particle states as a function of crystal momentum $k$.
  The periodic potential $H_{\rm pot}$ couples states with the same
  $k$, splitting the degeneracies, giving rise to the band structure
  shown as a dotted grey line.  In b) scattering processes are
  illustrated.  On the left hand side a generic scattering event is
  shown where two particles with arbitrary momenta, scatter to two
  other states.  On the right a `vertical' scattering event is shown,
  where two particles with the same crystal momentum scatter to two
  other states, preserving $k$.  These vertical scatterings are
  included in $H_{\rm int}^{\rm (vert)}$, while all others are in
  $H_{\rm int}^{\prime}$}
\label{hammeaning}
\end{figure}

The meaning of each of these terms is illustrated in
FIG.~\ref{hammeaning}.  Solid lines show the kinetic energy of free
particles as a
function of the crystal momentum $k$.    The periodic potential
conserves the crystal momentum and therefore only induces vertical
transitions.  The main effect of $H_{\rm pot}$ is therefore to split
the degeneracies at the level crossings, giving rise to the band
structure shown in gray.    The two interaction terms scatter
particles between these states.  The `vertical' interaction $H_{\rm
  int}^{\rm (vert)}$, only involves particles that all share the same
crystal momentum, as illustrated on the right of
 FIG.\ref{hammeaning}b.  All other
scattering processes, such as the one on the left of
 FIG.\ref{hammeaning}b are
included in $H_{\rm in}^\prime$.

In our subsequent analysis we will ignore $H_{\rm in}^\prime$.  This
is a quite drastic approximation which clearly restricts the phenomena which can be
studied.  For example, the superfluid-insulator transition seen in
\cite{mott} cannot be studied in this model.  However, 
all studies of atoms in
optical lattices which rely upon mean field theory (the
Gross-Pitaevskii equation) implicitly make this truncation whenever
they limit themselves to a Bloch ansatz \cite{wunote}.  This
approximation therefore has a range of validity which 
is a superset of the mean field theory's.  In particular this
approximation is good when the interaction strength is the smallest
energy in the problem.

Once $H_{\rm in}^\prime$ is eliminated, the sectors of different $k$
are independent.  If one identifies $k$ with the $\Phi$ in
(\ref{modham}), then the two Hamiltonians are identical.  For the
remainder of the paper, we work with (\ref{modham}), while keeping in
mind that all results can also be applied to a gas of particles in a
periodic lattice. 

\section{Energy landscape in absence of impurity}
Here, and in the next section, we calculate the properties of the
microscopic model (\ref{modham}), finding the general structures
discussed in the introductory sections.  We divide the discussion into
several sections, based upon the limits of various parameters and the
mathematical techniques used.
\subsection{two-mode approximation}\label{twomode}
\begin{figure}
\begin{tabular}{ll}
\parbox[b]{0.04\columnwidth}{a)\\[1in]}&\includegraphics[width=0.95\columnwidth]{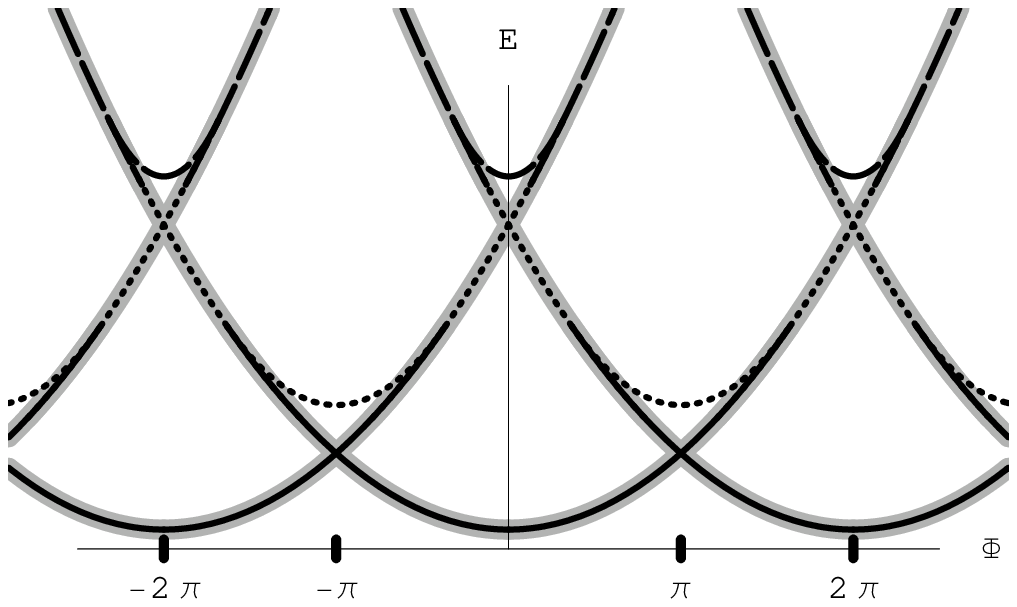}\\
\parbox[b]{0.04\columnwidth}{b)\\[1in]}&
\includegraphics[width=0.95\columnwidth]{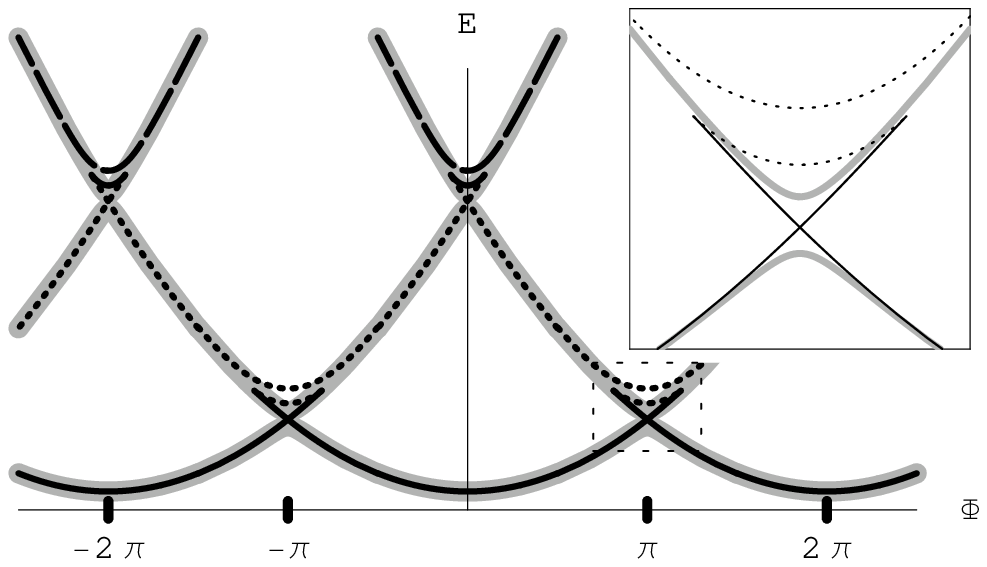}
\end{tabular}
\caption{Spectra:
Thick grey lines lines -- single particle energy levels of
  non-interacting particles in
  a 1-D ring of length $L$.  Black 
  lines -- mean-field energy extrema
  of interacting 
  system, with global mean-field shift removed.  Solid
  are local minima, dotted (dashed) are saddle points/cusps 
  with one (two) direction of
  negative curvature/slope.  Figures a) and b) are respectively with and
  without an added impurity.  In b) the area around $\Phi=\pi$ is
  enlarged and displayed in an inset.
 Notice the similarities between
  the single particle states for particles in the ring shown here, 
  and the band structure for particles in a periodic potential in
  FIG.~\ref{hammeaning} }
\label{energies}
\end{figure}
\begin{figure}
\includegraphics[width=0.9\columnwidth]{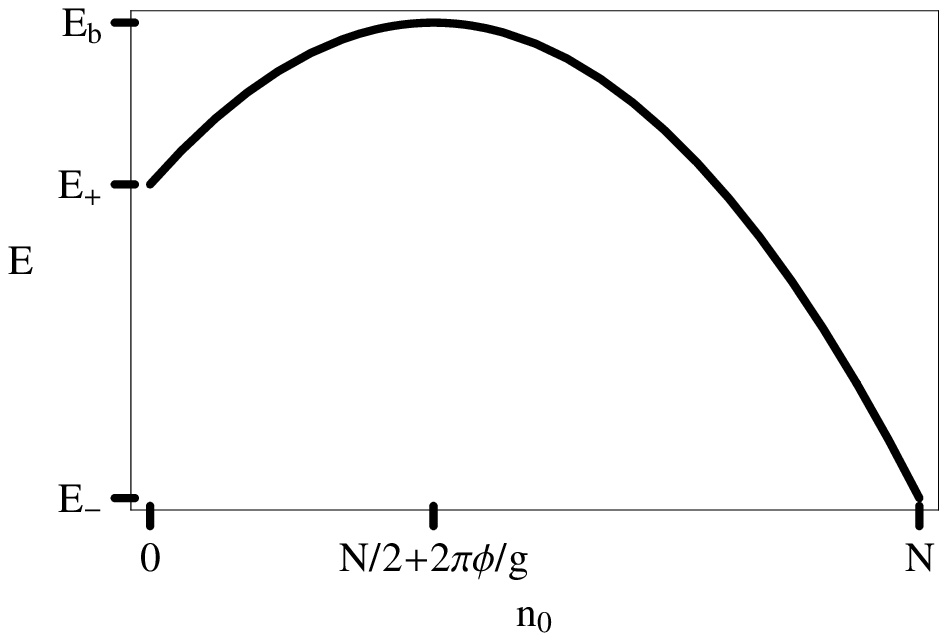}
\caption{Energy barrier separating persistent current carrying states
  within the two-mode approximation.  The ordinate shows $n_0$, the number of
  particles in the $l=0$ state, the remaining $N-n_0$ particles are in
  the $l=1$ state. The maximum occurs at 
  $n_0=\bar n_0=N/2+2\pi\phi/g$, where
  $\phi$ is a measure of the rotation speed, and $g$ the interaction
  strength.  A barrier only exists when $0<\bar n_0>N$.
  The labeled energies $E_b$ and $E_{\pm}$ are given in the
  text.}\label{barrier}
\end{figure}

It is instructive to first analyze (\ref{modham}) in the absence of an
impurity, and in the limit where the interactions are sufficiently
weak, ie. where $\lambda=0$ and $gN\ll1$, 
with $N$ being the number of particles.  The non-interacting
single-particle energy states are shown as thick grey lines 
in FIG.~\ref{energies} as a function
of $\Phi$.  This spectrum and the physical properties which we are
interested in are periodic in $\Phi$, and it suffices to
consider $-\pi \leq\Phi\leq \pi$. 
The ground state, in the absence of interactions,
 consists of all particles condensed in the lowest
energy state.  Aside from providing a global shift in the chemical
potential, weak interactions only introduce a significant
perturbation when the
energy difference between two
levels is less than $gN/2$, so that the interactions mix the two
states.  Focusing on the level crossing at $\Phi=\pi$, the system
is reduced to two-levels with an effective Hamiltonian
\begin{equation}\label{twolevel}
H_\pi = (\phi+\pi)^2 n_0 + (\phi-\pi)^2 n_1 + 
\frac{g}{2} (n_0^2+n_1^2+4 n_0 n_1),
\end{equation}
where $\phi=\Phi-\pi$, and $n_j=c_j^\dagger c_j$ are constants of
motion.  The eigenstates 
$|n_0,n_1\rangle=(c_0^\dagger)^{n_0}
(c_1^\dagger)^{n_1}|0\rangle/\sqrt{n_0! n_1!}$, have a fixed number of
particles in each momentum state.  Just like in the noninteracting
system, for $\phi<0$ (respectively
$\phi>0$) the ground state is $|N,0\rangle$ ($|0,N\rangle$).
Interactions play a role here only through the fact that when  
$\gamma=2\pi |\phi|/g<N/2$ 
a barrier, as illustrated in FIG.~\ref{barrier}, separates
these two states.  This barrier exists because 
density modulations are required 
to transfer
particles between angular momentum states.
In the presence of interactions these
modulations cost energy.  The barrier state,
$|N/2+\gamma,N/2-\gamma\rangle$, has energy 
$E_b=(\phi+\pi)^2(N/2+\gamma)+(\phi-\pi)^2(N/2-\gamma)+
3gN^2/4-g\gamma^2$, compared with
$E_{\pm}=(\phi\pm\pi)^2N+gN^2/2$ for the other extrema.  The maximum barrier
height (occurring at $\phi=0$) is $\delta E=gN^2/4$.

This same scenario is repeated at all other level crossings in
FIG.~\ref{energies}.  Thus, in this weakly interacting limit,
one can take the eigenstates of EQ. (\ref{modham}) to be the `Fock' states
\begin{equation}
|n_0,n_1,n_{-1}\cdots\rangle = 
\prod_j \frac{\left(c_j^\dagger\right)^{n_j}}{\sqrt{n_j!}} |0\rangle,
\end{equation}
where the occupation numbers $n_j$ obey the constraints $n_j>0$ and
$\sum_j n_j=N$.  For large numbers of particles there is no
approximation involved in thinking of the $n_j$ as continuous variables.

We have already detailed the energy landscape when we
truncate this space to two $n_j$'s, and the space of allowed states
consists of a line (the x-axis of the plot in FIG.~\ref{barrier}).  
When three $n_j$'s are included, the space is a
triangle, and with four $n_j$'s it is a tetrahedron.  The
d-dimensional generalization of a triangle/tetrahedron is 
often called
a simplex or a hyper-triangle, 
and within our approximations, the eigenstates of the
Hamiltonian form an infinite dimensional simplex.  The corners of
this simplex are cusps in the energy topography.  Cusps play a role
similar to saddle points, as they are each classified by the number of
independent directions in which the energy decreases.  

As explicitly
shown in FIG.~\ref{barrier} for
the case of two levels, there exists a range of $\Phi$ for which one can find
extrema in addition to these cusps.  In 
FIG.~\ref{energies}, the extrema are marked by the number of
`downward' directions.  Loops are clearly visible around points where
the non-interacting states cross.  These loops get smaller when the
interaction strength decreases.  Taking $g$ and $\Phi$ as the control
parameters, one reproduces the structure in FIG.~\ref{threedswallow}.






As in our generic discussion, these loops lead to hysteresis.
Suppose 
our model 1-D gas starts in
the state $|N,0\rangle$ at $\Phi=0$, 
and is then accelerated so that $\Phi$ is slightly
larger than $\pi$.  The barrier will then
keep the system from jumping
into the new ground state $|0,N\rangle$.  Thus the presence of this
barrier implies the existence of persistent currents.  
Of course, since we have
considered only very small interactions, these currents only
exist near $\Phi=\pm\pi$.

\subsection{Mean field theory}\label{sec:mf}
A useful tool to further illucidate our model (\ref{modham})
 is mean field theory.  Here we use a mean-field theory to show that the
general structure of metastability and superfluidity
 found within the two-mode approximation continues to 
be valid for larger interaction strengths.

This discuss reveals several important points.
(1) Even though a one-dimensional Bose gas 
is a Luttinger liquid and
 is usually not studied using mean field theory, we show below that
{\em  mean field theory correctly
 describes the behavior of a one-dimensional Bose gas for a
 significant parameter range}.  The exact details of this parameter range 
is discussed below.  (2)  The extrema of the mean-field Hamiltonian 
are in one-to-one correspondence with the energy extrema 
in the many-body Hilbert space discussed in section~\ref{twomode}.
(3) In the regime where both the two-mode approximation and mean-field theory are applicable, 
the Bogoliubov excitation spectrum of the mean-field coincides with the exact low energy
excitations of the many-body problem.  This correspondence is well-known from the Bethe ansatz analysis of the one-dimensional Bose gas \cite{lieb}.

\subsubsection{The two-mode regime}
We begin by considering mean field theory in the regime 
where the two-mode approximation is valid.
For weakly interacting Bosons, mean field theory
can be formulated as a variational method in which one assumes all of
the particles are in the same single particle state.  In the
effective two-level Hilbert space described by the Hamiltonian in 
EQ. (\ref{twolevel}), we
 consider wavefunctions of the form $|\alpha,\beta\rangle =
(\alpha a_0^\dagger +\beta a_1^\dagger )^N |0\rangle$, where the
variational parameters $\alpha$ and $\beta$ satisfy
$|\alpha|^2+|\beta|^2=1$, and $|0\rangle$ is the vacuum state
containing no particles.  For $\phi<0$ ($\phi>0$), the energy is
minimized by $|\alpha|^2=1$, $|\beta|^2=0$
($|\alpha|^2=0,|\beta|^2=1)$, which (within the two-mode approximation)
is the exact ground state as shown
in SEC.~\ref{twomode}.
When $|\phi|< g(N-1)/4\pi$, one also finds a mean-field state which is
a maximum of the energy, corresponding to the barrier state found
previously.  The topography of the mean-field energy landscape mirrors
 that of the exact eigenstates, and the barriers found within the
 mean field theory do not differ significantly from the exact barriers.

It is quite striking that the mean field theory compares so favorably with the 
exact two-mode calculation,
considering that the exact barrier state contains two, rather than one,
condensates and is therefore referred to as
``fragmented'' \cite{frag}.  The connection between these states
is understood by noting that
that the mean-field barrier corresponds to a `phase slip', where
the fluid density vanishes at some point.  By its nature such an event
breaks rotational symmetry.  Averaging over the possible location of
the slip restores the broken symmetry, and leads to the exact
(fragmented) barrier state \cite{muellerfrag}.

In this weakly interacting limit
the excitations of the mean field theory
correspond to the exact low-lying excited states of the two-mode 
system.   This result is trivially obtained by substituting our variational 
wavefunction into (\ref{twolevel}) and calculating the frequencies of small oscillations.
For those familiar with dilute gases of bosons, this result is perhaps even simpler to
derive
by going beyond the two mode approximation, and 
directly writing down the excitations
of a condensate moving at velocity $v_c$.
As with a three-dimensional system, an excitation of wave-vector $k$ 
is given by the Bogoliubov form,
\begin{equation}\label{bog}
E_k=\frac{\hbar^2}{2m}\sqrt{k^2\left(k^2
+2 g N/L^2\right)}-v_c k
\end{equation}
Here the finite size of the ring restricts the wavevector to $k=2\pi
n/L$, with integer $n$.  Similarly $v_c=2\pi\hbar n^\prime/(m L)$ is
quantized with integer $n^\prime$.  
In the frame rotating with velocity $\Omega$ the
excitations have energies $E_k-\hbar\Omega L/2\pi$.
Within the two-mode approximation we are limited to $n=1$, and
it is straightforward to verify that 
excitation spectrum matches the low energy spectrum calculated
directly from (\ref{twolevel}). 
Exploring the excitation spectrum around the saddle-point state, one
finds a zero-mode corresponding to translations of the phase
slip, and negative energy modes corresponding to falling towards one
of the local minima.   In the exact two-mode theory, the zero mode 
corresponds to changing the relative phase between the modes.

\subsubsection{Beyond two-modes}
In addition to the insights provided above,
the mean-field approach also provides a systematic way to 
explore EQ.~(\ref{modham}) for interaction strengths 
which are beyond the scope of
the two mode approximation.
The mean-field theory involves replacing the field operators
$c_k$
with c-numbers.  It is convenient to work in real space, defining
a 'condensate wavefunction' $\psi(x)$ by
\begin{equation}
c_k = \int_0^1 {\rm d}x\,e^{ikx}\psi(x),
\end{equation}
where $\int{\rm d}x|\psi(x)|^2=N$, is the number of particles.
An energy landscape can be found in the space of all possible 
square integrable complex-valued functions $\psi(x)$.
This landscape contains all of the structures seen in the 
two-mode version of mean field theory.  In particular the $k=n$ states given
by
$\psi_n(x)=\sqrt{N} e^{i 2 n x}$, are always stationary points.  
Their stability is given by the Landau criterion that if
the excitation spectrum in 
EQ.~(\ref{bog}) is positive then they are local minima.  
Otherwise there exists a direction of negative curvature.  
The existence of multiple
minima in the energy landscape leads to hysteresis and superfluidity.

When both the $k=n$ and $k=n-1$ states are locally stable,
there must exists a saddle point separating them.  As in the two-mode case
this saddle point
involves a 'phase slip', where the density vanishes and the number of units of circulation
can change. 
The real space wavefunction of the 
phase slip
takes on the form of a hyperbolic
trigonometric function, whose exact form was determined 
by Langer and Ambegaokar in the context of superconductors \cite{langer}.  
The barrier height is understood by
recognizing that the length-scale for the phase slip is the healing
length $\xi$ ($\approx1/\sqrt{gN}$ in our dimensionless units).  The
presence of the slip increases the density from $N$ to $N/(1-\xi)$,
at an energy cost per particle of 
$g N \xi\to\hbar^2\sqrt{a_s L/d_\perp^2}/mL^2$ in physical
units (assuming transverse harmonic confinement with lengthscale
$d_\perp=\sqrt{\hbar/m\omega}$).  A more careful calculation, detailed in
\cite{mueller}, verifies this result with a coefficient $\sqrt{32/9}$.  

\subsubsection{limits of validity}
We have shown that within mean-field theory the one-dimensional Bose gas
is superfluid in that it exhibits hysteresis under changing the rotation speed.
It is therefore very important to understand the limits of validity of mean-field theory.
We estimate these limits 
by calculating the depletion
within a Bogoliubov approach (for example see \cite{Fetter}), where one finds that
\begin{equation}\label{deltan}
\frac{\delta N}{N} =\sum_{k} \frac{\epsilon_k-E_k}{2 E_k}
\sim \sqrt{\frac{g}{N}}\log(g N),
\end{equation}
where $\epsilon_k=\hbar^2/2m(k^2+ g N/L^2)$. 
Only when this ratio is small compared to $1$ is the mean-field theory
valid.  For experimentally relevant parameters, the logarithm is of
order ten, and the prefactor
$\sqrt{g/N}$ determines the size of the depletion.
A physical interpretation of this factor
is that the healing length $\xi$ ($1/\sqrt{gN}$)
which governs the scale of the
phase slip must be larger than the
interparticle spacing ($1/N$) for fluctuations to be small and for
mean-field theory to be applicable
(ie. phase slips cost very little energy if they fit between
particles).
To study stronger interactions one needs to include
short-range fluctuations -- either through `bosonizing' the system
\cite{haldane}, or by using the Bethe ansatz \cite{lieb}.

Substituting plausible experimental values into EQ.~\ref{deltan}, show
that it is much easier to be in the regime where mean field theory is
applicable than it is to be in the strongly correlated regime.  For
example, with $10^6$ atoms of $^{85}$Rb (with scattering length
$a_s\approx 5$nm) in a ring of circumference
$L=100\mu$m, and transverse confinement 
$\omega_\perp=500 s^{-1}$
(corresponding to $d_\perp\sim 1 \mu$m) one finds $\delta N/N \sim
0.2\%$.  Decreasing $N$ or increasing $L$ leads to proportionally
more depletion and can bring one into the strongly correlated regime.

\section{persistent currents in presence of impurity}\label{imp}
Having established the superfluid behavior of a weakly interacting
gas through analysis of the energy landscape, we now analyze the behavior 
of such a system when a small impurity is added.  We shall see that as
long as the impurity strength is small compared to the interactions, such an impurity
leads to extremely slow (typically exponentially slow) decay of persistent currents.
In the opposite limit, where the impurity is strong, no persistent currents nor hysteresis exists.

Even when the impurity is weak, it does have a dramatic effect on the
(mean-field) energy landscape in FIG.~\ref{energies}, in that it opens up
gaps as seen in FIG.~\ref{energies}b).   In the case of
atoms in a periodic potential these
gaps are the familiar band-gaps from solid-state physics discussed in the introduction.  Note that
the gaps do not change the fact that one has hysteretic behavior signaled by the 
swallow-tail loops.  In this section we present the quantitative theory of these loops in the limit of weak interactions and calculate the quantum tunneling from one local minimum to another (ie. the decay of persistent currents induced by quantum fluctuations).

\subsection{two-mode model}
Interactions are most easily understood within a two-mode
model.  For its validity we will need both weak interactions $gN\ll1$,
and a weak impurity $\lambda\ll1$.  The ratio $\lambda/gN$ which
compares the impurity strength to the interaction strength is
arbitrary.  As before, it suffices to consider the system near
$\phi=\Phi-\pi=0$, where the Hamiltonian may be truncated to (\ref{twolevel}),
with an additional impurity term $H_{\rm imp} = \lambda (a_1^\dagger
a_0 + a_0^\dagger a_1)$.  This system of $N$ bosons in two states can be
mapped onto the precession of a spin $N/2$ object obeying a Hamiltonian
\begin{equation}\label{spn}
H_{\rm spin}=4\pi\phi S_z+ 2\lambda S_x-2 g S_z^2+C_N
\end{equation}
where $S_z=( a_0^\dagger a_0-a_1^\dagger a_1)/2$,
$S_x=(S_++S_-)/2$, $S_y=(S_+-S_-)/2i$,
$S_+=a_0^\dagger a_1$, and $S_-=a_1^\dagger
a_0$ obey the standard spin algebra, and $C_N$ is an uninteresting
c-number.
This mapping is the inverse of the method of ``Schwinger
bosons'' \cite{schwingerbose}.  
The quantum dynamics of such large spins are well
understood \cite{bigspin}, so we only briefly outline the analysis necessary 
to understand the mean-field structure, and the decay of persistent
currents.

\begin{figure}[tbph]
\vspace{0.1in}
\begin{tabular}{|ll|ll|}
\hline
\parbox[b]{0.01\columnwidth}{a)\\[1in]}&
\includegraphics[width=0.46\columnwidth]{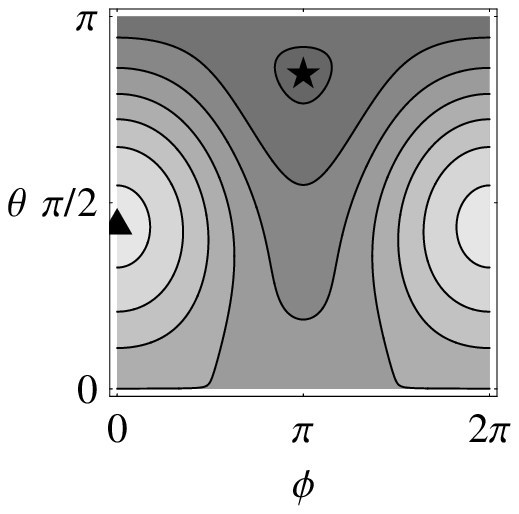}&
\parbox[b]{0.01\columnwidth}{b)\\[1in]}&
\includegraphics[width=0.46\columnwidth]{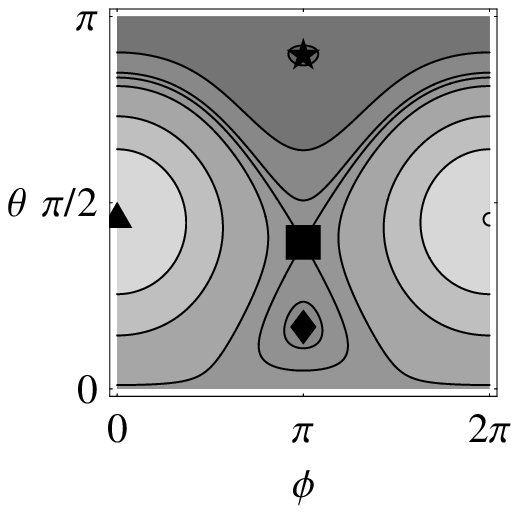}\\
\hline
\parbox[b]{0.01\columnwidth}{c)\\[0.5in]}&\includegraphics[width=0.46\columnwidth]{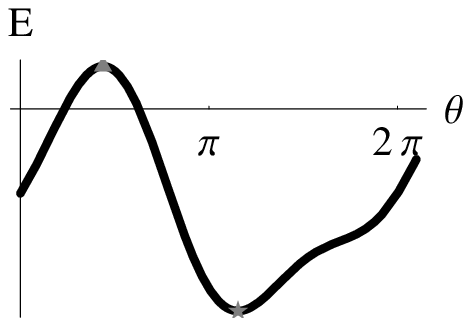}&
\parbox[b]{0.01\columnwidth}{d)\\[0.5in]}&
\includegraphics[width=0.46\columnwidth]{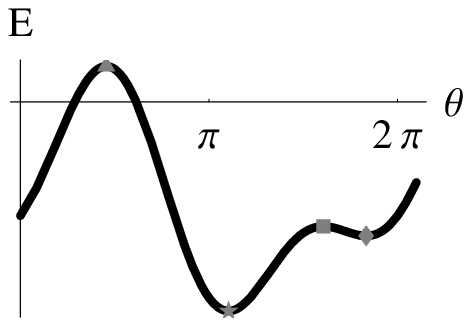}\\
\hline
\end{tabular}
\caption{
Energy landscape of spin which represents the state of 
the one dimensional Bose gas 
within a two-mode approximation in the presence of an
impurity.
  The number of
  particles with angular momentum $l=0$ ($l=\hbar$) 
  is $N \cos^2(\theta)$ ($N
  \sin^2(\theta)$), while $\phi$ is the phase angle describing the
  coherence between these states.  For the contour plots, a) and b), 
  darker colors represent lower
  energy; stars, triangles, diamonds, and squares represent the global
  minimum, global maximum, local minimum and saddles.  In a)  the
  inequality in EQ.~(\ref{condition}) is not satisfied, and only two
  extrema exist, while in
  b) the inequality is satisfied.  A nonlinear scale is used for the
  contours in b) to emphasize the extrema.  All of the extrema occur
  on a great circle parameterized by setting $\phi=0$ and letting
  $\theta$ run from $0$ to $2\pi$.  This corresponds to a path
  from the `south pole' to the `north pole'
  along the `meridian' where $\phi=0$, and returning along the
  $\phi=\pi$.  Figures~c) and d) show the energies of a) and b) as a
  function of $\theta$ along these paths.  The structures from
  FIG.~\ref{genlandscape} are clearly seen.}  \label{landscape}
\end{figure}

The classical (mean field) energy landscape of the spin is shown in
FIG.~\ref{landscape}, where the angles $\theta$ and $\phi$ describe the direction
in which the `spin' is pointing.  In the absence of interactions, there are two stationary
points, a maximum and a minimum.  These represent the first and second
band in the single-particle energy spectrum.  For sufficiently strong
interactions
\begin{equation}\label{condition}
(gN)^{2/3}>(2\pi\phi)^{2/3}+\lambda^{2/3},
\end{equation}
a second minimum and a saddle spontaneously
appear (a saddle-node bifurcation), 
giving rise to the swallow-tail energy spectrum shown in
FIG.~\ref{energies}.   The new minimum is analogous to the metastable states
found in the absence of the impurity.  The difference here is that
this metastable state is distinct from the upper band, which is still
represented by the maximum.  
In the symmetric case ($\phi=0$), the minima occur at
$S_z=N/2\sqrt{1-(\lambda/gN)^2}$, and the barrier has a height
$E_b=gN^2/2 (1-\lambda/gN)^2$
as is clear from
(\ref{condition}), the metastable state can only exist if the
interaction strength $gN$ exceeds the impurity strength $\lambda$.

\subsection{Quantum tunneling in the two-mode limit}
Quantum mechanically, there are matrix elements for tunneling from the
upper minima to the lower, and the upper state acquires a finite
lifetime.   For those more familiar with particle
tunneling than with spin tunneling it may be helpful to instead map
the problem onto the motion of a particle on a one-dimensional lattice.
Introducing states $|m\rangle=(-1)^m|n_0=N/2+m,n_1=N/2-m\rangle$ and
operators $c_m^\dagger$, which create these states, the many-body
problem in the two-mode approximation is equivalent to a single
particle with a Hamiltonian
\begin{widetext}
\begin{equation}\label{tunham}
H=\sum_m \left[\left(4\pi\phi m -g m^2\right) c_m^\dagger c_m
- \lambda
\sqrt{\left(\frac{N}{2}+m+1\right)\left(\frac{N}{2}-m\right)}
\left(c_{m+1}^\dagger c_m+c_m^\dagger c_{m+1}\right) \right]+C_N^\prime,
\end{equation}
\end{widetext}
where, $C_N^\prime$ is an uninteresting c-number.  This
Hamiltonian represents a particle on a lattice with an inverted
parabolic potential and an unusual spatially dependent hopping (which
can be viewed as a
spatially dependent mass).  A particle surmounting a barrier by
hopping on such a discrete
lattice behaves somewhat differently than a similar particle with a
continuous coordinate.  In particular, for sufficiently large
barriers, the tunneling rate is a power-law in $\lambda$ rather than
the familiar exponential.
Such power-law behavior was predicted by
Kagan et al. in discussing the finite temperature lifetime of
persistent currents \cite{kagan}.
\begin{figure}[tbph]
\includegraphics[width=0.8\columnwidth]{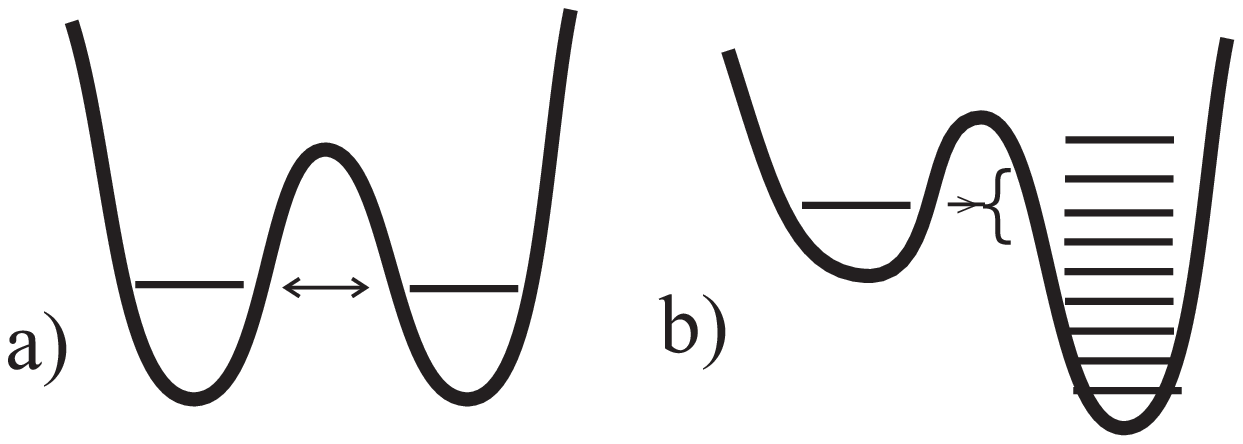}
\caption{Schematic of tunneling within a double well system: a)
  coherent tunneling between resonant states;  b) incoherent decay
  from a single state on the left to a large number of states on the
  right.  (Here this decay corresponds to a transition from a current carrying
  state to a stationary one with a large number of phonon
  excitations.)
}\label{wells}
\end{figure}

An important conceptual point to consider here is to what extent
the scenario discussed so far can lead to the decay of a current.  We
have reduced the many-body problem to the one-dimensional quantum
mechanical motion of a single particle.  There is no source of
dissipation within this model, and one would naively
expect to see coherent
oscillations between the two wells rather than a decay.  This
intuition is in fact correct when the two wells are degenerate
(ie. $\phi=0$), or when the tunneling is extremely weak.  This
coherent tunneling limit is
illustrated in FIG.~\ref{wells}a, in which only  the
lowest state in each well is coupled.
When there is a significant
mismatch in the energies of the two wells, say the left hand well has
more energy than the other, the situation is different.  As
pictured in FIG.~\ref{wells}b, the ground
state on the left can be coupled to several excited states on the right, a
situation analogous to an excited atom coupled to a large number of
vacuum modes.  
In the limit of large $N$, the spacing between the
modes in the right hand well vanish, and the state on the left hand
side of the barrier acquires a finite lifetime.  
A detailed study of
this cross-over from coherent oscillations to decay is found in
\cite{benderskii}.  One expects decay whenever the tunneling rate is
significantly larger than $\Delta/\hbar$, where $\Delta$ is the
characteristic energy spacing 
(here given by the inverse of the density of states of Bogoliubov phonons).

A semiclassical analysis of (\ref{tunham}) is
detailed in \cite{bigspin}.  The eigenvalues of (\ref{tunham}) are
found by solving a difference equation.  This difference equation can
be approximated by a differential equation which is amenable to a WKB
analysis.  The resulting expression for the lifetime $\tau$
of the persistent
current is
\begin{equation}\label{cosh}
\tau = \tau_0 \exp\left(
\int_{b_1}^{b_2} \!\! {\rm arccosh}
\frac{-2g s^2+4\pi\phi s-E}{2\lambda\sqrt{(N/2)^2-s^2}}\,\,\,
{\rm d}s\right)
\end{equation}
The attempt frequency $\tau_0^{-1}$ is roughly the
frequency of small oscillations about the local minimum of (\ref{spn}).
This frequency
also coincides with the quantum mechanical energy of excitations
in the metastable state
(which as already pointed out, is the energy of Bogoliubov
excitations).  For the symmetric case, $\phi=0$, 
one finds 
$\tau_0^{-1}\sim 2 \sqrt{(gN)^2-\lambda^2}$.
The integration variable $s$ corresponds to the
projection of spin $S_z$, the
limits of integration $b_1$ and $b_2$ are the
classical turning points, and $E$ is the energy, given by (\ref{spn})
with the constant $C_N$ removed.  When
$s=b_{1,2}$ the argument of the ${\rm arccosh}$ is $1$.

\begin{figure}
\vspace{0.25in}
\psfrag{logrt}[bc]{\small $\log\left( \tau^{-1}\,\, {\rm sec.}\right)$}
\psfrag{lam}[cl]{\small $\lambda/\hbar\omega$}
\includegraphics[width=0.8\columnwidth]{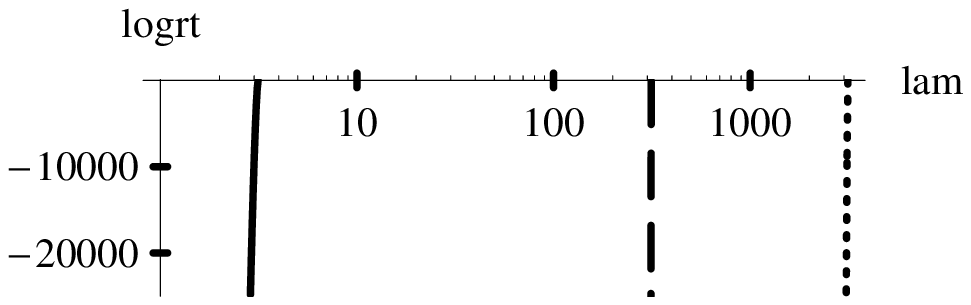}
 \caption{
 Lifetime of a metastable current carrying state (\ref{cosh}) in a
 toroidally trapped Bose gas in the presence of an impurity of strength
 $\lambda$. For this figure, we
 consider the case where the trap is rotating at exactly one half
 quantum of circulation $\phi=0$.  In the curves shown, the trap is
 described by a circumference $L$, a transverse confinement frequency
 $\omega$, a number of particles $N$.
These
 are respectively: solid line, $L=10^{-2}$m, $N=10^6$, 
$\omega=500 s^{-1}$;
dashed line, $L=10^{-4}$m, $N=10^7$, $\omega=500 s^{-1}$;
dotted line, $L=10^{-2}$m, $N=10^8$, $\omega=50000 s^{-1}$.  
 One expects
 that imperfections in the apparatus would 
 lead to $\lambda\ll \hbar\omega$.  
In all cases the particles are taken to have scattering length
$a_s=5$nm and mass $m=$85
 atomic units.
The metastability turns on so
 quickly that the curves appear nearly vertical, even on this
 logarithmic scale.
 }
 \label{rate}
 \end{figure}

For sufficiently small barriers, $E_b\ll \lambda N$, the $\rm arccosh$
can be expanded as ${\rm arccosh}(1+x)= \sqrt{2 x}+{\cal O}(x^{3/2})$,
  yielding a decay rate which is exponentially small in the impurity
  strength, $\tau^{-1}\sim \exp(- \alpha \sqrt{N E_b/\lambda})$, where
  $\alpha$ is of order unity.  For larger barriers, one can expand
${\rm arccosh}(z)= \log(2 z) + {\cal O}(z^{-2})$,  which leads to a
power law behavior, $\tau^{-1} \sim (N \lambda/E_b)^{-\beta E_b/N}$, where
$\beta\approx 1$.  Unless the barrier is tuned extremely close to
zero, both of these expressions yield astronomically large lifetimes
whenever a barrier exists.  Thus, the condition for superfluidity
reduces to the condition for a barrier (\ref{condition}).
 In FIG.~\ref{rate},
 equation (\ref{cosh}) is numerically integrated for some
 representative parameters, verifying these asymptotics.

\section{acknowledgments}

This work is
supported by NASA Grants NAG8-1441, NAG8-1765, and by NSF Grants
DMR-0109255, DMR-0071630.
The author would like to thank Tin-lun Ho for his kind support and
insightful suggestions, and both Q. Niu and 
Mehmet Oktel for critical comments.  
The use of a two-mode approximation to study
superfluidity was first introduced to me by A. J. Leggett in private
communications (the two-mode model with 
$g<0$ is discussed in \cite{uedaleggett}).


\begin{thebibliography}{99}
\bibitem{leggett} A. J. Leggett, Rev. Mod. Phys. {\bf 73}, 307 (2001).

\bibitem{Niu}B. Wu and Q. Niu, Phys. Rev. A {\bf 61}, 023402 (2000).

\bibitem{pethick} Dmitri Diakonov, L. M. Jensen, C. J. Pethick, and
  H. Smith, cond-mat/0111303 (2001).

\bibitem{carr} Much of the study of swallowtails is 
rooted in the exact solutions to the Gross-Pitaevskii equation derived in;
J.~C. Bronski, L.~D. Carr, B.~Deconinck, and J.~N. Kutz,
Phys. Rev. Lett. {\bf 86} 1402 (2001);
 J.~C. Bronski, L.~D. Carr, B.~Deconinck, J.~N. Kutz,
and K. Promislow, 
Phys. Rev. E {\bf 63} 036612 (2001);
 J.~C. Bronski, L.~D. Carr, 
R. Carretero-Gonz\'alez,
B.~Deconinck, J.~N. Kutz,
and K. Promislow, 
Phys. Rev. E {\bf 64} 056615 (2001).

\bibitem{landau}L. D. Landau and E. M. Lifshitz, {\em Statistical
    Physics Part 1} (Pergamon, New York, 1980).

\bibitem{thom} R. Thom, {\em Structural Stability and Morphogenesis}
  (Benjamin, Reading, Massachusetts, 1975).

\bibitem{hepersist}  The hypothetical experiment discussed here is an
  amalgam of W. F. Vinen Proc. Roy. Soc. {\bf A260}, 218 (1961),
  S. C. Whitmore and W. Zimmermann Phys. Rev. {\bf 166}, 181 (1968),
  and P. W. Karn, D. R. Starks, and W. Zimmermann, Phys. Rev. B {\bf
    21}, 2797 (1980).

\bibitem{leggett2} See A. J. Leggett, J. Stat. Phys. {\bf 93}, 927
  (1998) or Rev. Mod. Phys. {\bf 71}, S318 (1999) for a detailed discussion.

\bibitem{feynman} R. P. Feynman, {\em Statistical Mechanics,}
  (Addison-Wesley, Reading 1972) chapter 11.

\bibitem{opticallattices} Mark Raizen, 
Christophe Salomon and Qian Niu, Phys. Today, {\bf 50}, 30 (1997).

\bibitem{niuold} D. Choi and Q. Niu, Phys. Rev. Lett. {\bf 82}, 2022 (1999).

\bibitem{mott} 
Markus Greiner, Olaf Mandel, Tilman Esslinger, Theodor W. H\"ansch,
and Imamanuel Bloch, {\em Nature} {\bf 415}, 39 (2002).

\bibitem{ashcroft} Neil W. Ashcroft and N. David Mermin, {\em Solid
    State Physics,} (Sanders College, Fort Worth 1976).

\bibitem{niulz} Biao Wu, R. B.  Diener, 
and Qian Niu, Phys. Rev. A {\bf 65}, 025601 (2002).

\bibitem{mitinterfere} M. R. Andrews, D. M.
  Kurn, H.-J. Miesner, D. S.  Durfee, C. G. Townsend, S. Inouye, and
  W. Ketterle, Phys. Rev. Lett.  {\bf 79}, 553 (1997), {\em ibid.}
  {\bf 80}, 2967 (1998).

\bibitem{jjexample} A lucid discussion can be found in \cite{leggett} pp 336.  Also see references therein.

\bibitem{chapman}J. A. Sauer, M. D. Barrett, and M. S. Chapman,
  Phys. Rev. Lett. {\bf 87}, 270401 (2001).

\bibitem{haldane} F. D. M. Haldane, Phys. Rev. Lett. {\bf 47} 1840 (1981).

\bibitem{lieb} Elliot H. Lieb and Werner Liniger, Phys. Rev. {\bf
    130}, 1605 (1963);  Elliot H. Lieb, Phys. Rev. {\bf 130}, 1616 (1963).

\bibitem{kagan} Yu. Kagan, N. V. Prokof'ev, B. V. Svistunov,
  Phys. Rev. A {\bf 61}, 045601 (2000).

\bibitem{wunote} The mean-field studies of Wu et al. \cite{Niu,niuold}, relax the Bloch ansatz to calculate the stability of the mean-field state.  Their analysis can be reproduced by perturbatively including $H^\prime_{\rm int}$.

\bibitem{frag} P. Nozieres and D. Saint James, J. Phys. (Paris) {\bf
    43}, 1133 (1982).

\bibitem{muellerfrag} Erich J. Mueller, Gordon Baym, Tin-Lun Ho, and
  Masahito Ueda, {\em in preparation}.


\bibitem{langer} J. Langer and V. Ambegaokar, Phys. Rev. {\bf 164}, 498 (1967).

\bibitem{mueller} Erich J. Mueller, Paul M. Goldbart, and Yuli
  Lyanda-Geller, Phys. Rev. A {\bf 57}, R1505 (1998).

\bibitem{Fetter} Alexander L. Fetter and John D. Walecka, {\em Quantum theory
of many-particle systems}  (McGraw-Hill, San Francisco, 1971).

\bibitem{schwingerbose} see J. J. Sakurai {\em Modern Quantum
    Mechanics} (Addison-Wessley, Reading, 1994) p217.

\bibitem{bigspin} J. L. Van Hemmen and A. S\"ut\"o, Physica {\bf
    141B}, 37 (1986).

\bibitem{benderskii} V. A. Benderskii, and E. I. Kats, Phys. Rev. E
  {\bf 65}, 036217 (2001).



\bibitem{uedaleggett} Masahito Ueda and Anthony J. Leggett,
  Phys. Rev. Lett. {\bf 83}, 1489 (1999).

\end{thebibliography}
\end{document}